\documentstyle[twocolumn,aps]{revtex}

\begin{document}

\draft

\title{
\hfill {\rm UCSBTH--95--38~~~and~~~cond-mat/9512115} \\
$\phantom{X}$ \\
Comment on ``Spatial Correlation in Quantum Chaotic Systems with
Time-Reversal Symmetry: Theory and Experiment''}

\author{Mark Srednicki}

\address{Department of Physics, University of California,
         Santa Barbara, CA 93106}

\maketitle

\begin{abstract}
\normalsize{Abstract:
We show that the results of Prigodin et al can be reproduced and simplified
by making use of Berry's conjecture that the energy eigenfunctions in a
quantized chaotic system are gaussian random variables.
}
\end{abstract}


Prigodin et al \cite{prig2} have computed the joint probability distribution
\begin{equation}
P(v_1,v_2) =
\langle\delta(v_1-V|\psi({\bf x})|^2)\delta(v_2-V|\psi({\bf y})|^2)\rangle
\label{probdef}
\end{equation}
for the squared amplitude of an energy eigenfunction $\psi({\bf x})$
at two different points ($\bf x$ and $\bf y$) in a quantum dot with
volume $V$, assuming unbroken time-reversal invarance.
This was accomplished by averaging over a random
potential using supermatrix techniques.  The result, their Eq.\ (5),
is expressed as a parametric double integral.  However, this result
is equivalent to the considerably simpler expression
\begin{eqnarray}
P(v_1,v_2) &=& {1\over 2\pi(1-f^2)^{1/2}(v_1 v_2)^{1/2}}
                \exp\left(-\,{v_1+v_2\over 2(1-f^2)}\right) \nonumber \\
\noalign{\smallskip}
           && \times\cosh\left({f\sqrt{v_1 v_2}\over 1-f^2}\right).
\label{probv}
\end{eqnarray}
Here $f=f(|{\bf x}-{\bf y}|)=V\langle\psi^*({\bf x})\psi({\bf y})\rangle$,
and the angle brackets denote averaging over the random potential.
We will not need the explicit formula for $f(r)$, given in Eq.\ (7)
of \cite{prig2}, but recall that $f(0)=1$.

Eq.\ (\ref{probv}) can be derived from the conjecture
(originally due to Berry \cite{berry}) that the energy eigenfunctions in a
quantized chaotic system are (for sufficiently high energy eigenvalues)
Gaussian random variables.  We interpret this to mean that
\begin{equation}
P(\psi) \propto \exp\left[-\,{\beta\over2}\int d{\bf x}\,d{\bf y}\,
                    \psi^*({\bf x})K({\bf x},{\bf y})\psi({\bf y})\right],
\label{probpsi}
\end{equation}
where $P(\psi)$ is the probability that a particular
energy eigenfunction (with a definite energy eigenvalue)
is equal to the specified function $\psi({\bf x})$.
Here $\beta=1$ for a system which is time-reversal invariant,
and $\beta=2$ for a system which is not.  In the former case
(the one considered in \cite{prig2}), $\psi({\bf x})$ is a real function.
In either case, the kernel $K({\bf x},{\bf y})$ is the inverse of the
two-point correlation function
$\langle\psi^*({\bf x})\psi({\bf y})\rangle=V^{-1}f(|{\bf x}-{\bf y}|)$,
where the angle brackets now denote an average over $P(\psi)$.
We note that an assumption equivalent to Eq.\ (\ref{probpsi})
was made in \cite{alhas} to calculate the probability
distribution of level widths and conductance peaks in a
quantum dot with attached leads.

To get Eq.\ (\ref{probv}) from Eq.\ (\ref{probpsi}), we note that
integrating out all variables except
$\psi_1=\psi({\bf x})$ and $\psi_2=\psi({\bf y})$
will yield a Gaussian in these variables, and this Gaussian
must reproduce the correct two-point correlation functions.
Thus we conclude that
\begin{equation}
P(\psi_1,\psi_2) \propto (\det{M})^{-\beta/2}
                         \exp\left[-\,{\textstyle{\beta\over2}}
                                    \psi^*_i(M^{-1})_{ij}\psi_j\right],
\label{probpsi2}
\end{equation}
where
$M_{ij}=\langle\psi^*_i\psi_j\rangle = V^{-1}[\delta_{ij}+(1-\delta_{ij})f]$.
For the time-reversal invariant case, $\beta=1$ and $\psi_i$ is real;
changing integration variables to $v_i=V\psi_i^2$ and including a proper
Jacobian then yields Eq.\ (\ref{probv}).
Making use of standard combinatoric properties of Gaussian distributions,
it is straightforward to show that this expression for $P(v_1,v_2)$ has
the same moments as the one given in \cite{prig2}, and so must be, in fact,
the same function of $v_1$ and $v_2$.  Note, however, that the explicit
formula for the moments which is given in \cite{prig2}, their Eq.\ (21),
has a typographical error; the factor of $(2n-1)!!(2m-1)!!$ should be
$(2n)!(2m)!$ \cite{tanig}.

We can also reproduce the results of \cite{prig1} for the case of
broken time-reversal invariance.
Beginning with Eq.\ (\ref{probpsi2}) with $\beta=2$ and $\psi_i$ complex,
changing integration variables to $v_i=V|\psi_i|^2$
and $\theta_i=\arg\psi_i$,
including a proper Jacobian, and integrating over $\theta_1$
and $\theta_2$ yields
\begin{equation}
P(v_1,v_2) =
{1\over 1-f^2}\exp\left(-\,{v_1+v_2\over 1-f^2}\right)
              I_0\left({2f\sqrt{v_1 v_2} \over 1-f^2}\right),
\label{probnt}
\end{equation}
where $I_0(z)$ is a modified Bessel function.  This is
the same as Eq.\ (15) of \cite{prig1}.

\end{document}